\documentclass[10pt]{ismd08}
\usepackage{graphicx,epsfig}
\usepackage{cite,./mcite}
\usepackage{amsmath,amssymb}

\begin{document}
\title{Particle production and saturation at RHIC and LHC}
\author{Cyrille Marquet}
\institute{Institut de Physique Th\'eorique, CEA/Saclay, 91191 Gif-sur-Yvette cedex, France\\
Department of Physics, Columbia University, New York, NY 10027, USA\\
E-mail: cyrille@phys.columbia.edu}
\maketitle

\begin{abstract}

The Color Glass Condensate picture of the nuclear wave function at small-$x$ successfully predicted the suppressed production of high-$p_T$ particles at forward rapidities in deuteron-gold collisions at RHIC. This triggered more efforts which resulted in theoretical improvements and predictions for different observables which will provide further phenomenological tests. I review recent theoretical developments and discuss the resulting predictions.

\end{abstract}

\section{Saturation and the Color Glass Condensate}
\label{sec:1}

When probing small distances inside a hadron or nucleus with a hard process, one resolves their partonic constituents. Increasing the energy of the scattering process at a fixed momentum transfer allows to probe lower-energy partons, with smaller energy fraction $x.$ As the parton densities in the hadronic/nuclear wavefunction grow with decreasing $x,$ they eventually become so large that a non-linear (yet weakly-coupled) regime is reached, called saturation, where partons do not interact with the probe independently anymore, but rather behave coherently. 

The Color Glass Condensate (CGC) is an effective theory of QCD \cite{cgcrev} which aims at describing this part of the wave function. Rather than using a standard Fock-state decomposition, it is more efficient to describe it with collective degrees of freedom, more adapted to account for the collective behavior of the small-$x$ gluons. The CGC approach uses classical color fields: 
\begin{equation}
|h\rangle=|qqq\rangle+|qqqg\rangle+\dots+|qqqg\dots ggg\rangle+\dots\quad
\Rightarrow\quad|h\rangle=\int D{\cal A}\ \Phi_{x_A}[{\cal A}]\ |{\cal A}\rangle
\label{cgc}\ .\end{equation}
The long-lived, large-$x$ partons are represented by a strong color source $\rho\!\sim\!1/g_S$ which is static during the lifetime of the short-lived small-$x$ gluons, whose dynamics is described by the color field ${\cal A}\!\sim\!1/g_S.$ The arbitrary separation between the field and the source is denoted $x_A.$

The CGC wavefunction $\Phi_{x_A}[{\cal A}]$ is the fundamental object of this picture, it is mainly a non-perturbative quantity, but the $x_A$ evolution can be computed perturbatively. Requiring that observables are independent of the choice of $x_A,$ a functional renormalization group equation can be derived. In the leading-logarithmic approximation which resums powers of
$\alpha_S\ln(1/x_A),$ the JIMWLK equation describes the evolution of $|\Phi_{x_A}[{\cal A}]|^2$ with $x_A.$ The information contained in the wavefunction, on gluon number and gluon correlations, can be expressed in terms of n-point correlators, probed in scattering processes. These correlators consist of Wilson lines averaged with the CGC wavefunction, and resum powers of
$g_S{\cal A}.$

\begin{figure}[t]
\begin{center}
\epsfig{file=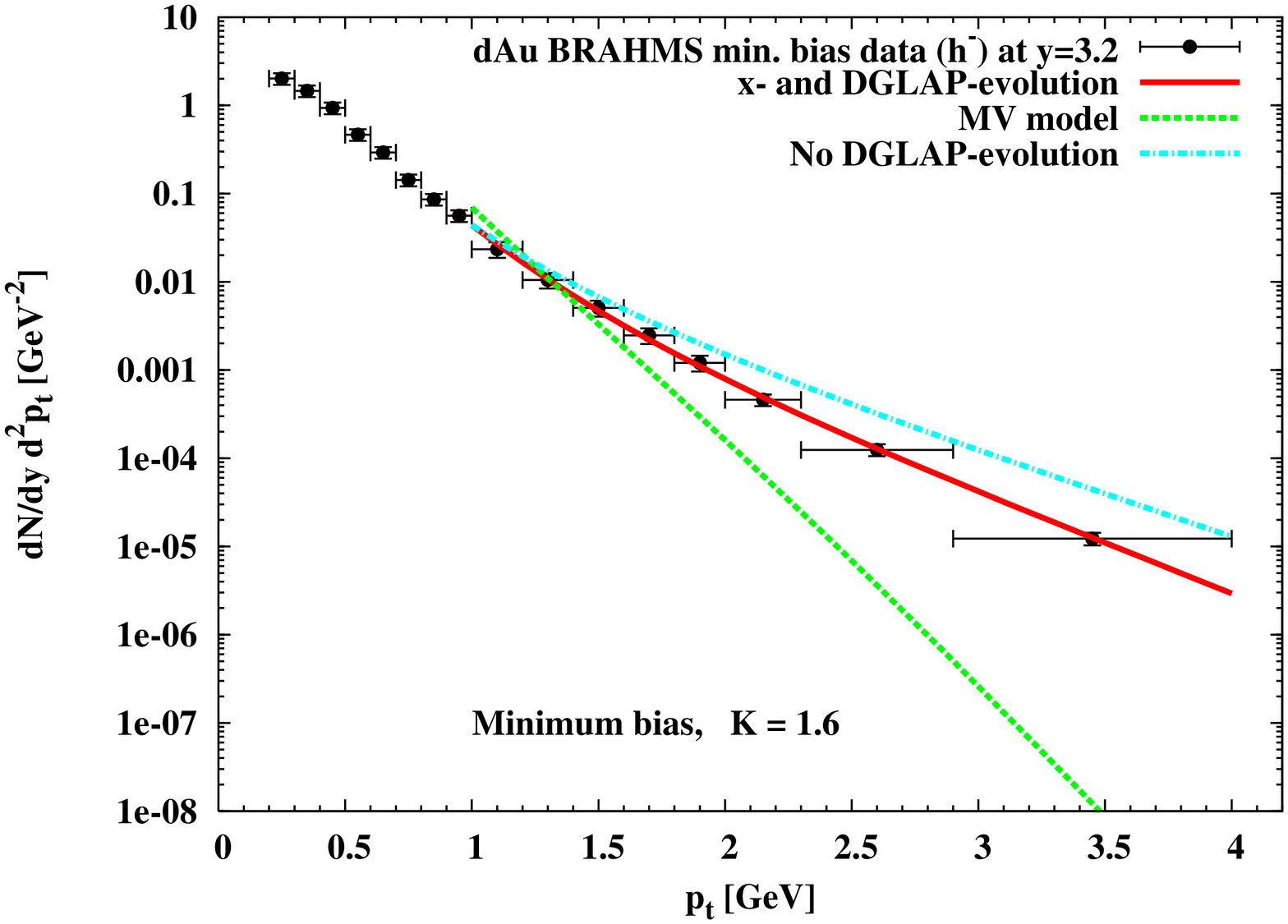,width=4.5cm,angle=-90}
\hspace{1cm}
\epsfig{file=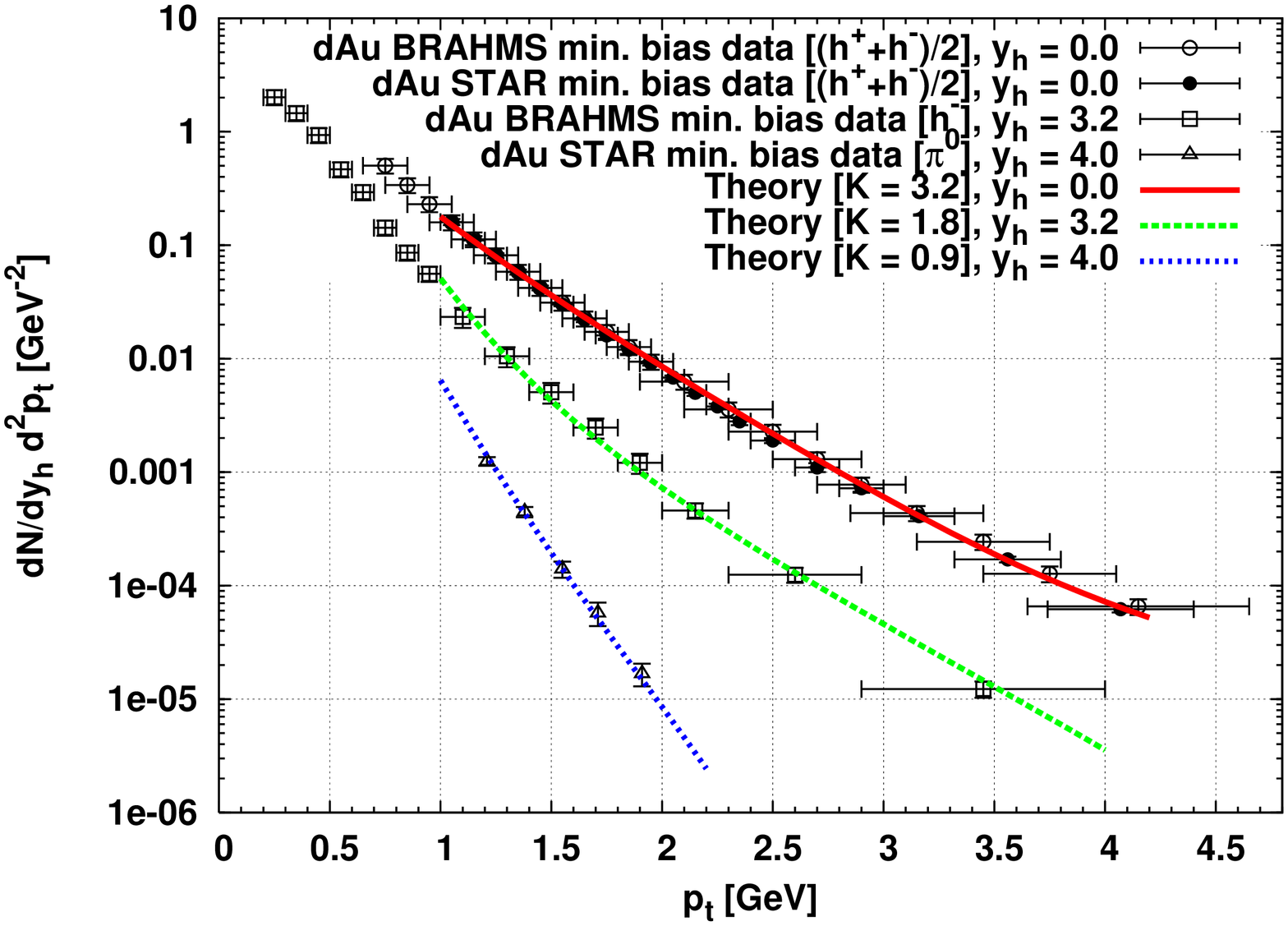,width=4.5cm,angle=-90}
\caption{Forward particle production in d+Au collisions at RHIC. The left plot shows the importance of including both the large-$x$ DGLAP evolution of the dilute deuteron and the
small-$x$ CGC evolution of the dense nucleus. The right plots shows the excellent description of the spectra shapes, and the K factors needed to obtain the normalization.}
\end{center}
\end{figure}

The JIMWLK equation reduces to a hierarchy of equations for the correlators. Most of the phenomenology uses a mean-field approximation which significantly simplifies the high-energy QCD evolution: it reduces the hierarchy to a single closed non-linear equation for the two-point function $\langle S_{\textbf{xy}}\rangle_{x_A}\!=\!1\!-\!\langle T_{\textbf{xy}}\rangle_{x_A},$ the Balitsky-Kovchegov (BK) equation \cite{bk-b1,bk-k1,*bk-k2}. It reads
\begin{equation}
\frac{d\langle S_{\textbf{xy}}\rangle_x}{d\ln(1/x)}=\frac{\bar\alpha}{2\pi}
\int d^2\textbf{z}\ M_{\textbf{xyz}}
\Big(\langle S_{\textbf{xz}}\rangle_x\langle S_{\textbf{zy}}\rangle_x
-\langle S_{\textbf{xy}}\rangle_x\Big)\ ,\quad M_{\textbf{xyz}}=
\frac{(\textbf{x}-\textbf{y})^2}{(\textbf{x}-\textbf{z})^2(\textbf{z}-\textbf{y})^2}\ ,
\label{bk}\end{equation}
with $\bar\alpha\!=\!\alpha_S N_c/\pi.$ All the correlators can then be expressed in terms of the solution of this equation. 
Finally, the Fourier transform of the dipole correlator 
$\int d^2\textbf{x}d^2\textbf{y}\ e^{i\textbf{k}.(\textbf{x}\!-\!\textbf{y})}
\langle T_{\textbf{xy}}\rangle_{x_A}/(\textbf{x}\!-\!\textbf{y})^2$
is an (all-twist) unintegrated gluon density. It determines forward particle production 
\cite{mygprod}, while more exclusive final states involve more complicated correlators. Solving
eq.~\eqref{bk} reveals the existence of an intrisic momentum scale in the nuclear wave function: the saturation scale $Q_s(x)$ which characterizes the transition from the dilute regime $k>Qs$ to saturation regime $k<Q_s.$

One of the most important progress is the recent calculation of the next-to-leading evolution equation \cite{nlobk-b1,*nlobk-b2,nlobk-k}. Concerning how the running coupling should be included, two schemes have been proposed by Balitsky (B) and Kovchegov and Weigert (KW). The following substitution should be done in formula \eqref{bk}, with
$R^2(\textbf{x},\textbf{y},\textbf{z})$ given in \cite{nlobk-k}:
\begin{equation}
\mathop{\frac{\bar\alpha}{2\pi}M_{\textbf{xyz}}}_{\hspace{0.3cm}\displaystyle\downarrow B}
\stackrel{KW}{\rightarrow}\frac{N_c}{2\pi^2}\left[\frac{\alpha_s((\textbf{x}\!-\!\textbf{z})^2)}
{(\textbf{x}\!-\!\textbf{z})^2}\!+\!2\frac{\alpha_s((\textbf{x}\!-\!\textbf{z})^2)
\alpha_s((\textbf{z}\!-\!\textbf{y})^2)}{\alpha_s(R^2(\textbf{x},\textbf{y},\textbf{z}))}
\frac{(\textbf{x}\!-\!\textbf{z})\cdot(\textbf{z}\!-\!\textbf{y})}
{(\textbf{x}\!-\!\textbf{z})^2(\textbf{z}\!-\!\textbf{y})^2}
\!+\!\frac{\alpha_s((\textbf{z}\!-\!\textbf{y})^2)}{(\textbf{z}\!-\!\textbf{y})^2}\right]
\end{equation}
\vspace{-0.7cm}
\begin{equation}
\frac{N_c\ \alpha_s((\textbf{x}\!-\!\textbf{y})^2)}{2\pi^2}
\left[M_{\textbf{xyz}}\!+\!\frac{1}{(\textbf{x}\!-\!\textbf{z})^2}
\left(\frac{\alpha_s((\textbf{x}\!-\!\textbf{z})^2)}
{\alpha_s((\textbf{z}\!-\!\textbf{y})^2)}\!-\!1\right)\!+\!\frac{1}{(\textbf{z}\!-\!\textbf{y})^2}
\left(\frac{\alpha_s((\textbf{z}\!-\!\textbf{y})^2)}
{\alpha_s((\textbf{x}\!-\!\textbf{z})^2)}\!-\!1\right)\right]\ .
\end{equation}
At next-to-leading order, there remains a discrepancy between the linear part of the BK equation and the BFKL equation. Running coupling corrections to particle production have also been investigated \cite{nlogprod}. Another important recent theoretical development is the inclusion of Pomeron loops in the evolution \cite{revdion,*revgreg}, and the derivation of potential consequences at very high energies \cite{diffscal,plgprod}. Concerning phenomenology at present colliders, there was however no significant impact.

\begin{figure}[t]
\begin{center}
\epsfig{file=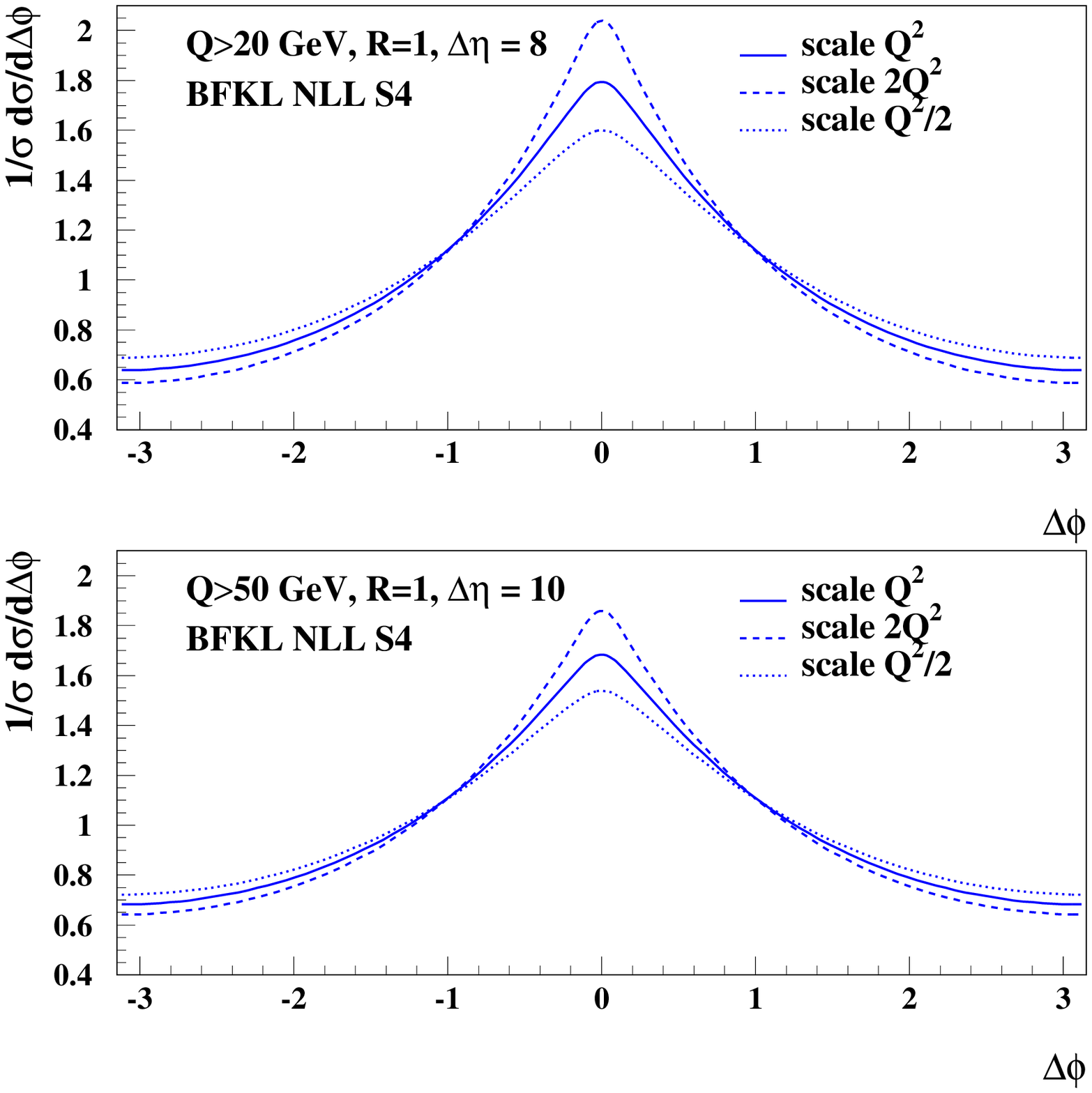,width=4.8cm}
\hspace{1cm}
\epsfig{file=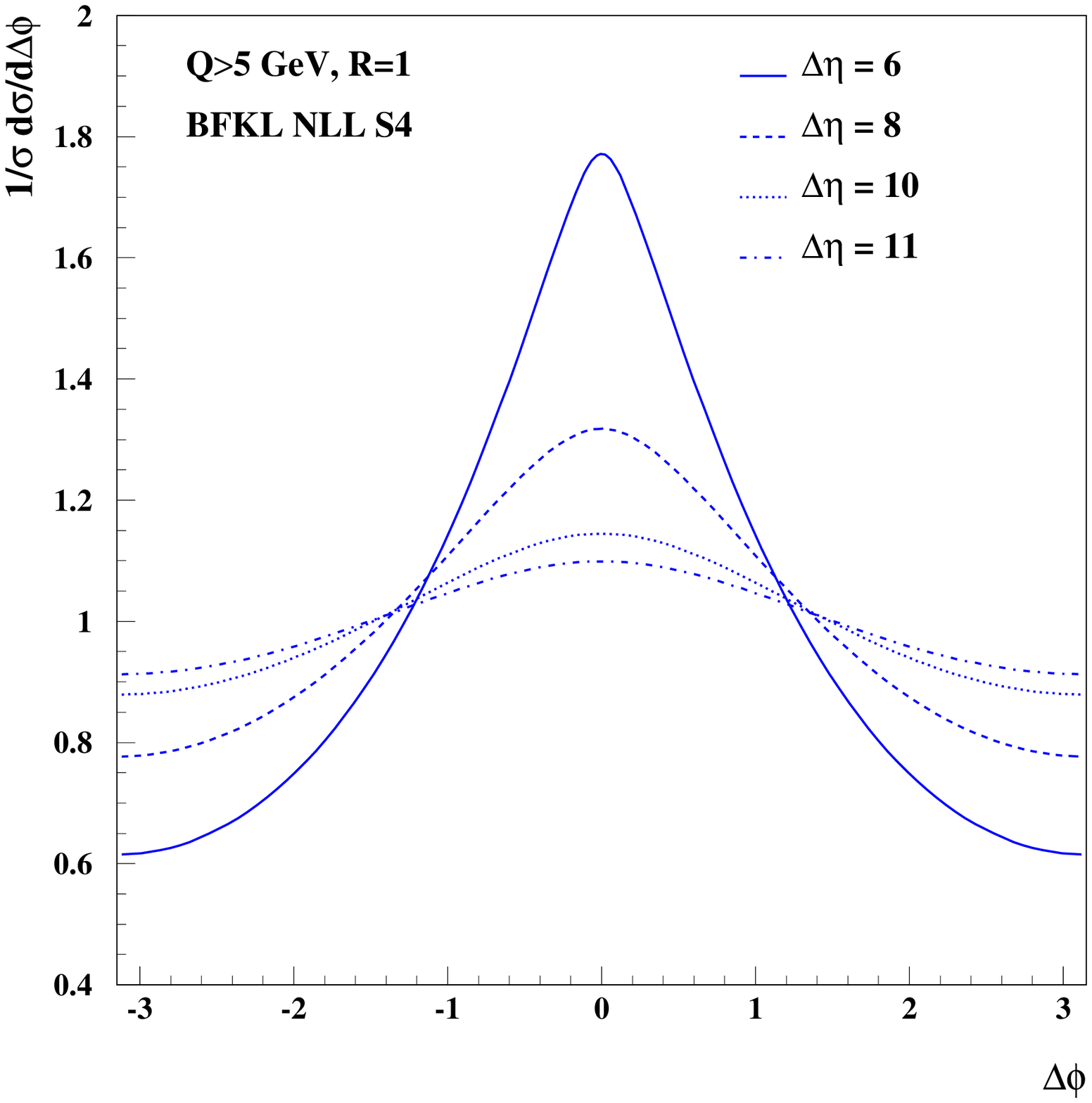,width=4.8cm}
\caption{Angular correlations of Mueller-Navelet jets in the NLL-BFKL framework.
Left plots: for standard Tevatron (top) and LHC (bottom) kinematics, the renormalization scale uncertainty is indicated (with $Q^2=k_1k_2$). Right plot: predictions for CDF (miniplugs) kinematics. The $\Delta\Phi$ distribution is peaked around $\Delta\Phi\!=\!0,$ which is indicative of jet emissions occuring back-to-back, and the $\Delta\Phi$ distribution flattens with increasing
$\Delta\eta$ or with $R$ deviating from 1.}
\end{center}
\end{figure}

\section{Forward particle production in pA collisions}
\label{sec:2}

Forward particle production in pA collisions allows to investigate the non linear QCD dynamics of high-energy nuclei with a probe well understood in QCD. Indeed, while such processes are probing small-momentum partons in the nuclear wavefunction, only high-momentum partons of the proton contribute to the scattering ($\sqrt{s} x_p\!=\!k e^y$ and $\sqrt{s} x_A\!=\!k e^{-y}$). The dilute hadron contributes via standard parton distribution functions while the CGC is described by its unintegrated gluon distribution. It was not obvious that the CGC picture \eqref{cgc}, which requires small values of $x_A,$ would be relevant at present energies. However, it has been the case for many observables in the context of HERA \cite{myrev} and RHIC \cite{jyrev}. One of the most acclaimed successes is the prediction that the yield of high-$p_T$ particles at forward rapidities in d+Au collisions is suppressed compared to A pp collisions, and should decrease when increasing the rapidity.

In Fig.1 the $dAu\!\to\!hX$ $p_T$ spectra computed in the CGC approach \cite{dhj1,*dhj2} is compared to RHIC data, and the description of the slope is impressive. The need of K factors to describe the normalization could be expected since this is a leading-order based calculation. Improving the calculation with the next-leading evolution has yet to be done. While the suppression was predicted in the CGC approach, other postdictions later offered alternative descriptions. The idea is that the value of $x$ probed in the deuteron is so high that large-$x$ effects could be responsible for the suppression \cite{dAualt-kop,*dAualt-fs}. This would not happen in $pA$ collisions at the LHC, with a smaller $x_p.$

While the CGC framework was quite successful in describing single inclusive particle production at forward rapidities, the focus should now shift towards more exclusive observables like two-particle production $pA\!\to\!h_1h_2X.$ In particular the correlations in azimuthal angle between the produced hadrons should be suppressed compared to pp collisions \cite{klm}. By contrast with single particle production, in two-particle production the CGC cannot be described only by its unintegrated gluon distribution, the so-called $k_T$-factorization framework is not applicable. This means that more tests could be done, probing the CGC structure deeper. The second d+Au run at RHIC gives the opportunity to carry out such measurements.

\begin{figure}[t]
\begin{center}
\epsfig{file=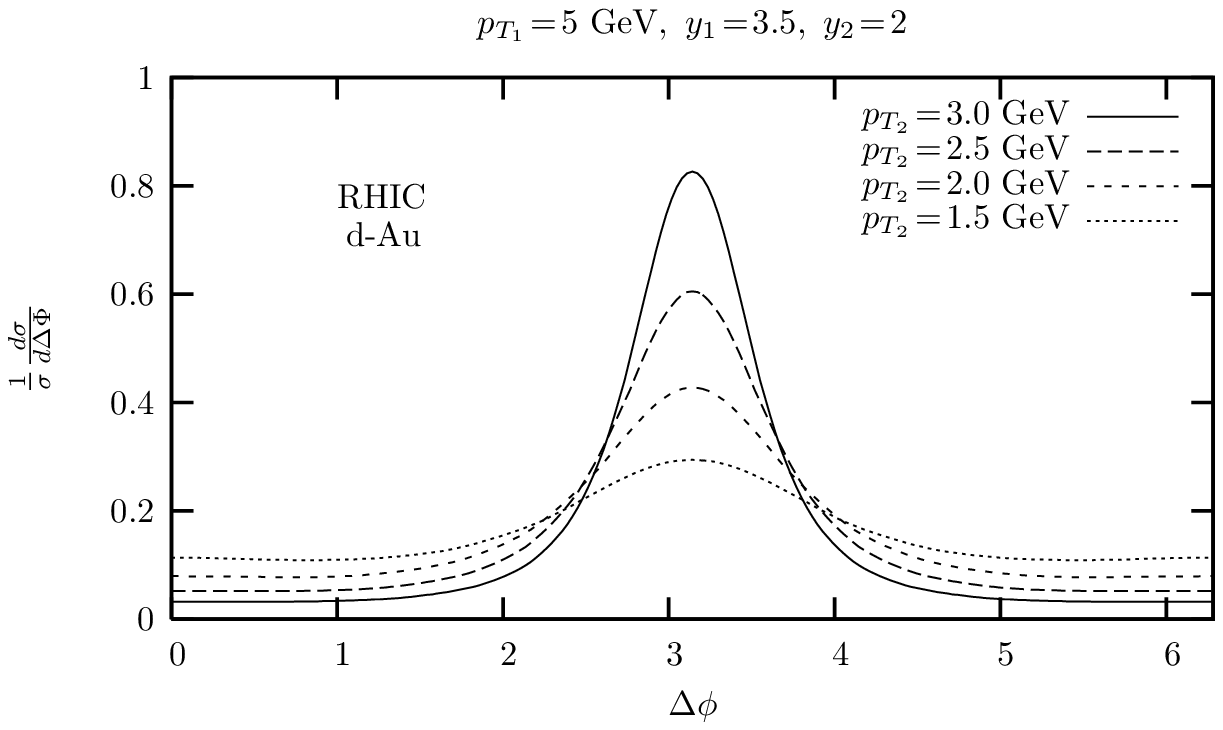,width=6.3cm}
\hspace{1cm}
\epsfig{file=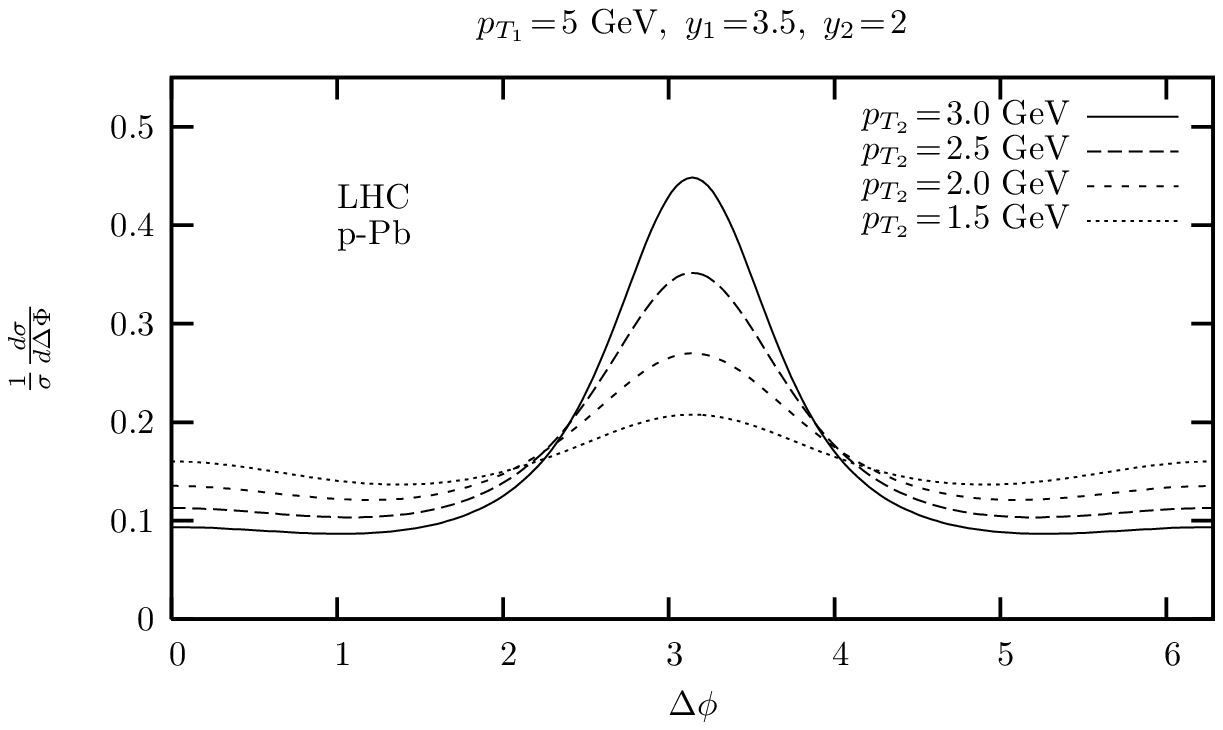,width=6.3cm}
\caption{Two-particle production at forward rapidities in pA collisions. The $\Delta\phi$ spectrum is displayed at RHIC (left) and LHC (right) energies. When varying $p_{T_2}$ at fixed $y_2,$ the correlation in azimuthal angle is suppressed as $p_{T_2}$ gets closer to the saturation scale. At the LHC, smaller values of $x_A$ are probed, and the azimuthal angle correlation is more suppressed as indicated by the vertical axis. The peak is also less pronounced.}
\end{center}
\end{figure}

\section{Selected predictions for the LHC}
\label{sec:3}

Mueller-Navelet jets \cite{mnjets} in hadron-hadron scattering are two jets produced in each of the forward directions. In the high-energy regime, in which the jets are separated by a large rapidity interval, this process is sensitive to the small-$x$ QCD evolution. 
An interesting observable is the azimuthal decorrelation of the jets as a function of their rapidity separation $\Delta\eta\!=\!y_1\!-\!y_2$ and of the ratio of their transverse momenta $R=k_2/k_1$ \cite{mnjnll-1,*mnjnll-2,*mnjnll-3}. Predictions are shown in Fig.2, for Tevatron and LHC kinematics, where $\Delta\Phi\!=\!\pi\!-\!\phi_1\!+\!\phi_2$ is the relative azimuthal angle between the two jets. The curves are obtained in the linear regime, using
next-to-leading logarithmic (NLL) BFKL evolution. At higher energies, saturation
effects will also be relevant \cite{mnjsat-1,*mnjsat-2,*mnjsat-3,*mnjsat-4}.

Coming back to forward particle production in pA collisions, predictions for the process
$pA\to h_1h_2X$ are shown in Fig.3, for RHIC and the LHC \cite{mytpc}. $k_1,$ $k_2$ and
$y_1,$ $y_2$ are the transverse momenta and rapidities of the final state hadrons, and the azimuthal angle spectra are displayed. It is obtained that the perturbative back-to-back peak of the azimuthal angle distribution (which is recovered for very large momenta) is reduced by initial state saturation effects. As the momenta decrease closer to the saturation scale ($Q_s\!\simeq\!2\ \mbox{GeV}$), the angular distribution broadens. But at RHIC energies, saturation does not lead to a complete disappearance of the back-to-back peak.

Finally, predictions for the total charged-particle multiplicity in AA collisions at the LHC are shown in Fig.4. Two approaches are compared: in the first, $k_T$-factorization is assumed but the evolution of the unintegrated gluon densities is accurately obtained from the next-leading BK equation \cite{javier}; in the second, the x evolution is only parametrized but multiple scatterings are correctly taken into account by solving classical Yang-mills equation
\cite{tuomas}. While full next-leading treatment of both multiple scatterings and small-$x$ evolution is desirable, the numbers obtained are similar, which indicates that the
uncertainties in both approaches are under control. 

\begin{figure}[t]
\begin{center}
\epsfig{file=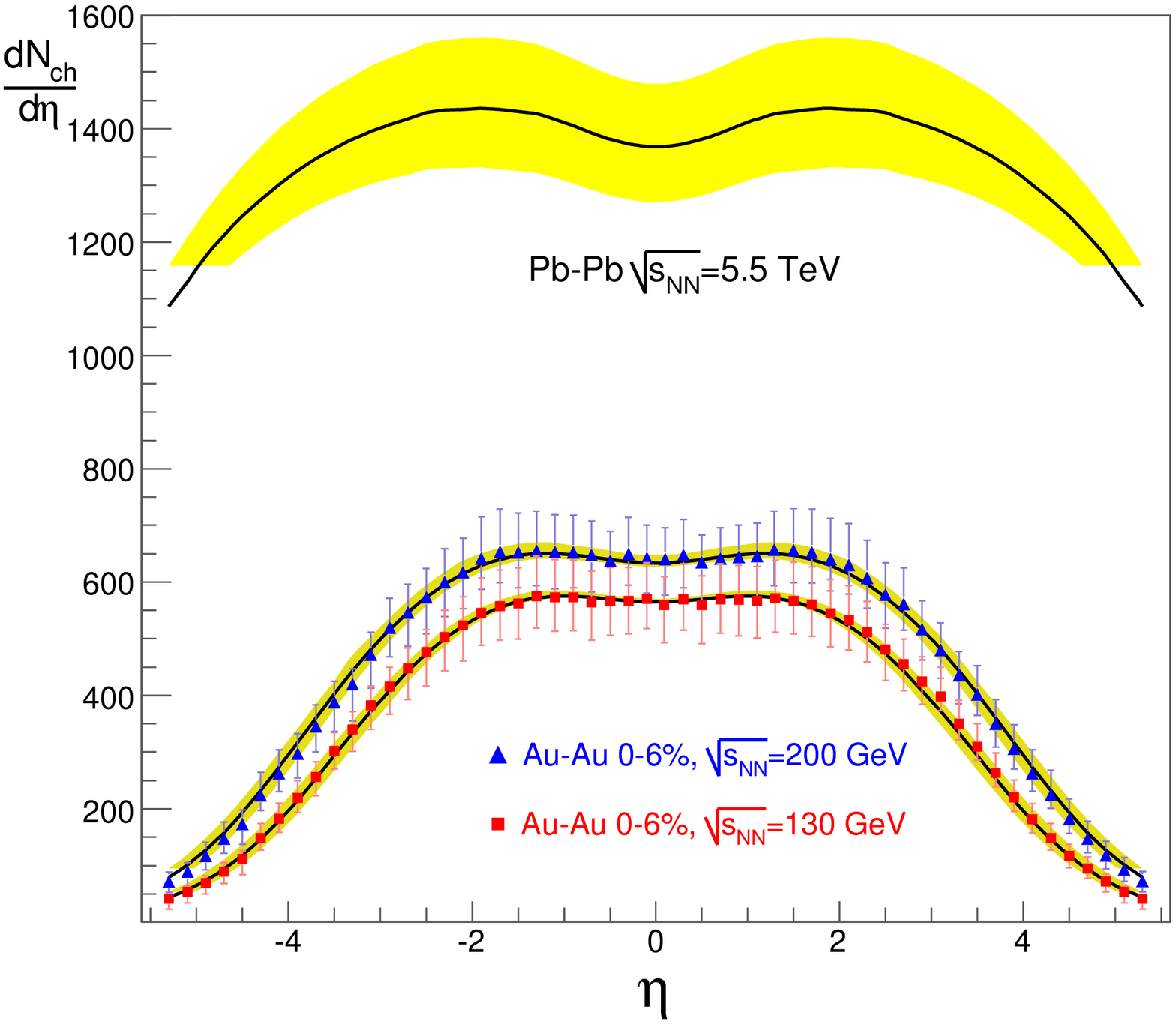,width=5cm}
\hspace{1cm}
\epsfig{file=tuomas.eps,width=6cm}
\caption{The charged-particle multiplicity in AA collisions at RHIC and the LHC. In both approaches a few parameters are fixed to reproduce RHIC data, such as the initial value of $Q_s.$ Then the small-$x$ evolution determines the multilicity at the LHC. The predictions are similar, around 1400 charged particles at mid rapidity for central collisions.}
\end{center}
\end{figure}

\begin{footnotesize}
\bibliographystyle{ismd08} 
{\raggedright
\bibliography{marquet}
}
\end{footnotesize}
\end{document}